\documentclass[reprint,amsmath, amssymb, aps, prd, superscriptaddress, nofootinbib, nobibnotes,longbibliography]{revtex4-1}

\usepackage{graphicx}
\graphicspath{{figs/}}
\usepackage{amsmath}
\usepackage{amssymb}
\usepackage{float}
\usepackage{color}
\usepackage[colorlinks=true, allcolors=blue]{hyperref}
\usepackage{acro}
\usepackage{ulem}

\DeclareAcronym{GR}{
	short = GR,
	long  = general relativity
	}
\DeclareAcronym{BH}{
	short = BH ,
	long  = black hole
}
\DeclareAcronym{BBH}{
	short = BBH ,
	long  = binary black hole
}
\DeclareAcronym{BNS}{
	short = BNS ,
	long  = binary neutron star
}
\DeclareAcronym{NSBH}{
	short = NSBH ,
	long  = neutron-star--black-hole
}
\DeclareAcronym{EFT}{
	short = EFT ,
	long  = effective field theory
}

\DeclareAcronym{GW}{
	short = GW ,
	long  = gravitational wave
}

\DeclareAcronym{CBC}{
	short = CBC,
	long  = compact binary coalescence
}

\DeclareAcronym{FRW}{
	short = FRW,
	long  = Friedmann-Robertson-Walker
}

\DeclareAcronym{PSD}{
	short = PSD,
	long  = power spectral density
}

\DeclareAcronym{SNR}{
	short = SNR,
	long  = signal-to-noise ratio
}

\DeclareAcronym{EM}{
	short = EM,
	long  = electromagnetic
}

\DeclareAcronym{SME}{
	short = SME,
	long  = standard model extension
}

\usepackage[capitalise]{cleveref}
\crefname{figure}{Fig.}{Figs.}
\Crefname{figure}{Fig.}{Figs.}

\def\be{\begin{equation}}
\def\ee{\end{equation}}
\def\({\left(}
\def\){\right)}
\def\[{\left[}
\def\]{\right]}

 \newcommand{\bqn}{\begin{eqnarray}}
 \newcommand{\eqn}{\end{eqnarray}}

\begin{document}
\title{Tests of Gravitational-Wave Birefringence with the Open Gravitational-Wave Catalog}

\author{Yi-Fan Wang}
\email{yifan.wang@aei.mpg.de}
\affiliation{Max-Planck-Institut f{\"u}r Gravitationsphysik (Albert-Einstein-Institut), D-30167 Hannover, Germany}
\affiliation{Leibniz Universit{\"a}t Hannover, D-30167 Hannover, Germany}

\author{Stephanie M. Brown}
\email{stephanie.brown@aei.mpg.de}
\affiliation{Max-Planck-Institut f{\"u}r Gravitationsphysik (Albert-Einstein-Institut), D-30167 Hannover, Germany}
\affiliation{Leibniz Universit{\"a}t Hannover, D-30167 Hannover, Germany}

\author{Lijing Shao}
\email{lshao@pku.edu.cn}
\affiliation{Kavli Institute for Astronomy and Astrophysics, Peking University, Beijing 100871, China}
\affiliation{National Astronomical Observatories, Chinese Academy of Sciences, Beijing 100012, China}

\author{Wen Zhao}
\email{wzhao7@ustc.edu.cn}
\affiliation{CAS Key Laboratory for Researches in Galaxies and Cosmology, Department of Astronomy, University of Science and Technology of China, Chinese Academy of Sciences, Hefei, Anhui 230026, China}
\affiliation
{School of Astronomy and Space Science, University of Science and Technology of China, Hefei 230026, China}

\begin{abstract}
We report the results of testing gravitational-wave birefringence using the largest population of gravitational-wave events currently available.
Gravitational-wave birefringence, which can arise from the effective field theory extension of general relativity, occurs when the parity symmetry is broken, causing the left- and right-handed polarizations to propagate following different equations of motion.
We perform Bayesian inference on the 94 events reported by the 4th-Open Gravitational-wave Catalog (4-OGC) using a parity-violating waveform.
We find no evidence for a violation of general relativity in the vast majority of events. However, the most massive event, GW190521, and the second most massive event, GW191109, show intriguing non-zero results for gravitational-wave birefringence.
We find that the probability of association between GW190521 and the possible \ac{EM} counterpart reported by Zwicky Transient Facility (ZTF) is increased when assuming birefringence.
Excluding GW190521 and GW191109, the parity-violating energy scale is constrained to $M_\mathrm{PV} > 0.05$ GeV at $90\%$ credible interval, which is an improvement over previous results from twelve events by a factor of five. 
We discuss the implications of our results on modified gravity and possible alternative explanations such as waveform systematics.
More detections of massive binary black hole mergers from the upcoming LIGO/Virgo/KAGRA run will shed light on the true origin of the apparent birefringence.

\end{abstract}

\maketitle
\section{Introduction}

The advanced LIGO and Virgo \cite{TheLIGOScientific:2014jea,TheVirgo:2014hva} detectors have completed three observation runs (O1-O3) and announced the detection of ninety confident \ac{GW} events in the Gravitational-Wave Transient Catalog (GWTC) \cite{GWTC-1,GWTC-2, GWTC-2-1, GWTC-3}.
Additional compact binary coalescence events are reported by independent analysis \cite{Nitz:2018imz,Nitz:2019hdf,Venumadhav:2019lyq,Nitz:2021uxj,Nitz:2021zwj,Olsen:2022pin} of the public data \cite{LIGO_Data}.
The most recent version of Open Gravitational-wave Catalog, 4-OGC \cite{Nitz:2021zwj}, used \texttt{PyCBC} toolkit \cite{pycbc-github} to search the public data from all three observation runs and reported ninety-four events \cite{Nitz:2021uxj}.  4-OGC  agrees with GWTC for all confident events with a probability of astrophysical origin $p_\mathrm{astro}>0.99$.
The detection of \ac{GW}s has enabled numerous precise tests of \ac{GR} in the strong, dynamical field \cite{LIGOScientific:2020tif, GWTC1-testingGR, GW150914testingGR, Chatziioannou_2021,PhysRevLett.123.191101, TestGr-GWTC3,Mehta2022} and high energy (sub-GeV) regimes \cite{Wang:2020cub}.
All the tests to date have confirmed that \ac{GW} data is consistent with the predictions of \ac{GR}.

This work tests GW birefringence using the currently largest population of gravitational-wave events from 4-OGC.
Birefringence of \ac{GW}s emerges when the parity symmetry, which is the invariance of physical laws regarding the inversion of spatial coordinates, is broken between the left- and right-handed \ac{GW} polarizations.
While parity symmetry is conserved  in \ac{GR}, theories where parity is violated have been proposed such as Chern-Simons gravity \cite{Yunes:2009yz,Yoshida:2017cjl,Chu:2020iil,Jung:2020aem,Kamada:2021kxi}, Ho\v{r}ava-Lifshitz gravity \cite{HL0,HL,Takahashi:2009wc}, ghost-free scalar-tensor gravity \cite{ghost}, and the symmetric teleparallel equivalent of \ac{GR} \cite{tele}  to account for dark matter and dark energy.
Parity violation also arises at high energy scales in quantum gravity theories such as loop quantum gravity and string theory \cite{Yunes:2009yz}.

We utilize an \ac{EFT} extension of the linearized Einstein-Hilbert action to study how deviations from \ac{GR} affect \ac{GW} propagation.
\ac{EFT} is a flexible framework that includes all action terms that purposely preserve or violate certain symmetries. 
The leading order higher derivative modification of the linearized action comes from terms of mass dimension five that violate parity \cite{prl2014,zhao2020}.
Parity violation leads to an asymmetry in the propagation speeds and amplitudes of the left- and right-hand polarization of \ac{GW}, which, in turn, leads to phase and amplitude birefringence, respectively.
Given the relationship between parity violation and Lorentz violation\cite{Greenberg:2002uu}, our tests have implications for constraining the \ac{SME} of gravity~\cite{Kostelecky:2003fs,Tasson:2016xib}, which is the most general \ac{EFT} extension of linearized \ac{GR} that violates Lorentz symmetry.

Tests of \ac{GW} birefringence were first done in ref.~\cite{mewes}, where they checked for waveform peak splitting in the first-ever detected \ac{GW} event, GW150914. 
Reference~\cite{1809} constrained birefringence using the \ac{GW} propagation speed measured from the binary neutron star merger event GW170817 \cite{GW170817}, and Refs.\cite{Nair:2019iur,Shao:2020shv,Wang:2020pgu,Yamada:2020zvt,Okounkova:2021xjv,Wang:2021ctl,Zhao:2022pun,Niu:2022yhr} placed further constraints on birefringence using GWTC events.

We test for \ac{GW} birefringence in ninety-four \ac{GW} events and find no evidence of birefringence in ninety-two events.
Intriguingly, we find two outliers, GW190521 and GW191109\_010717 (hereafter GW191109), the most and second most massive binary black holes in 4-OGC \cite{Abbott:2020tfl, Abbott:2020mjq,Nitz:2021zwj}, favor the birefringence waveform over the \ac{GR} waveform with Bayes factors $11.0$ and $22.9$, respectively ($\ln \mathcal{B}^\mathrm{nonGR}_\mathrm{GR} = 2.4$ and $3.1$).
Excluding these two outliers, we place constraints on the cutoff energy scale $M_\mathrm{PV} > 0.05$ GeV, which is more stringent than Ref.~\cite{Wang:2020cub} by a factor of five due to the larger data set and the more advanced waveform.
The result can be mapped to the \ac{SME} coefficient $\big| {\vec{\varsigma}}^{\,(5)} \big| < 9\times 10^{-16}\,{\rm m}$, which characterizes isotropic birefringence with mass dimension $d=5$.

The nonzero results may be caused by non-Gaussian noise fluctuation, systematic errors in waveform templates, or physical effects that are outside the standard assumptions for quasi-circular binary black hole mergers, e.g., the presence of orbital eccentricity \cite{Romero-Shaw:2020thy, Gayathri:2020coq}, a hyperbolic encounter \cite{Gamba:2021gap}, or even entirely new physics \cite{PhysRevLett.126.081101}.
We assess how the background noise affects the $M_\mathrm{pv}$ measurement for both GW190521 and GW191109 by injecting 100 \ac{GR} signals into the detector noise around the observed coalescence time for each event.  
The results suggest that our detection of \ac{GR} deviation for GW190521 and GW191109 is not likely an artifact of the noise.

Lastly, we consider the possible \ac{EM} counterpart of GW190521, which ZTF observed and comes from an active galactic nucleus \cite{Graham:2020gwr, Ashton:2020kyr}, by comparing the sky location posteriors for GW190521 with and without birefringence.
Prior analyses of GW190521 do not strongly favor association with the EM counterpart. 
Reference \cite{Nitz_2021} reported the Bayes factor of $\ln B \gtrsim -4$ which disfavors a sky location fixed to that of the electromagnetic counterpart.  
However, if we assume \ac{GW} birefringence, we find evidence in favor of the association with the \ac{EM} counterpart with $\ln \mathcal{B}$ of 6.6.

\section{Waveform templates for \ac{GW} birefringence}
\label{section:waveform}

We briefly overview the construction of waveform templates to test \ac{GW} birefringence following Ref.~\cite{zhao2020}. 
From the perspective of \ac{EFT}, the leading order modifications to the linearized Einstein-Hilbert action are terms with three derivatives and mass dimension five: $\epsilon^{ijk}\dot{h}_{il}\partial_{j}\dot{h}_{kl}$ and  $\epsilon^{ijk}\partial^2{h}_{il}\partial_{j}{h}_{kl}$ \cite{prl2014,zhao2020,Gao_2020}, where $i,j... = 1,2,3$ refer to spatial coordinates, $\partial_j$ denotes spatial derivatives, dot denotes derivatives with respect to time,  $\partial^2$ is the Laplacian, $\epsilon^{ijk}$ is the antisymmetric symbol, and $h_{ij}$ is the tensor perturbation of metric.  
Notably, both terms violate parity. 
Therefore, \ac{EFT} suggests that the leading order modification to \ac{GW} propagation arises from parity-violation.
We do not consider the more complicated anisotropic GW birefringence, for which the effects can be found in Refs.~\cite{mewes, ONeaLault2021analysis,Niu:2022yhr}.
Combining the above higher derivative terms with the linearized Einstein-Hilbert action gives
\begin{equation}
\label{eq:action}
\begin{split}
S = \frac{1}{16 \pi G}\int dtd^3x a^3\Bigg[\frac{1}{4}\dot{h}_{ij}^{2}- \frac{1}{4a^2}(\partial_k h_{ij})^2  + \\
\frac{1}{4}\( \frac{c_1}{aM_\mathrm{PV}}\epsilon^{ijk}\dot{h}_{il}\partial_j \dot{h}_{kl} + \frac{c_2}{a^3M_\mathrm{PV} }\epsilon^{ijk}\partial^2 h_{il} \partial_j h_{kl}\) \Bigg],
\end{split}
\end{equation}
where $a$ is the cosmic scale factor, $M_\mathrm{PV}$ is the energy scale at which higher order modification starts to be relevant, and $c_1$ and $c_2$ are two undetermined functions of cosmic time; 
the speed of light and the reduced Planck's constant are set to $c=\hbar=1$. 
Equation (\ref{eq:action}) is the generic form of the action; $c_1$, $c_2$ and $M_{\rm PV}$ can be mapped to the corresponding model parameters in a specific modified gravity theories \cite{Yunes:2009yz,Yoshida:2017cjl,Chu:2020iil,Jung:2020aem,Kamada:2021kxi,HL0,HL,Takahashi:2009wc,ghost,tele}, as explicitly demonstrated in ref. \cite{zhao2020}.

The equation of motion for the \ac{GW} circular polarization modes $h_A$, where $A = R$ or $L$ for the right- and left-hand modes, is
\be\label{eq:eom}
h_A'' +(2+\nu_A)\mathcal{H}h_A' + (1+\mu_A)k ^2 h_A = 0~,
\ee
where $\mathcal{H}$ is the conformal Hubble parameter, $k$ is the wavenumber, and a prime denotes the derivative with respect to the cosmic conformal time $\tau$, $\mu_A$ and $\nu_A$ are the phase and amplitude birefringence parameters.  They have the exact forms
\bqn\label{coes2}
\nu_{A}&=& - [\rho_{A}c_1k/(a M_{\rm PV})]'/\mathcal{H}, \nonumber \\ 
\mu_{A}&=&\rho_{A}(c_1-c_2)k/(a{M_{\rm PV}}),
\eqn
$\rho_{A}= \pm1$ for left- and right-handed polarizations represents broken parity of \ac{GW}s during propagation. 

We focus on velocity birefringence because the modification to \ac{GW} strain from amplitude birefringence is negligible \cite{zhao2020}. 
The \ac{GR} solution can be found by setting $\mu_A = \nu_A = 0$ in \cref{eq:eom}.

Solving \cref{eq:eom} gives the left- and right-handed circular polarization modes with parity violation. They are related to the \ac{GR} waveform by
\begin{equation}
\label{eq:pvwaveform}
h_L^{\rm PV}(f) = h_L^{\rm GR}(f)e^{-i\delta\Psi(f)}, ~h_R^{\rm PV}(f) = h_R^{\rm GR}(f)e^{i\delta\Psi(f)}.
\end{equation}
The plus ($h_{+}$) and cross ($h_{\times}$) modes of GW are given by $h_+=(h_L + h_R)/\sqrt{2}$ and $h_\times=(h_L-h_R)/(\sqrt{2}i)$.  

For binary black holes, we use the IMRPhenomXPHM \cite{Pratten:2020ceb} \ac{GR} waveform approximant, which includes sub-dominant harmonics and effects of precession; the IMRPhenomD\_NRTidal \citep{PhysRevD.93.044007, PhysRevD.93.044006, PhysRevD.96.121501, PhysRevD.99.024029} and IMRPhenomNSBH \citep{Thompson:2020nei}, waveforms, which account for tidal deformability, are used for binary neutron stars and neutron-star--black-hole mergers respectively. 

The phase modification to the \ac{GR} waveform takes the following form
\begin{equation}\label{eq:deltahpsi}
   \delta \Psi(f)=\frac{(\pi f)^2}{M_\mathrm{PV}} \int^{z}_{0} \frac{(c_1-c_2)(1+z')dz'}{H_0\sqrt{\Omega_M(1+z')^3 + \Omega_\Lambda}},
\end{equation}
where $H_0$ is the Hubble constant, $z$ is the cosmic redshift, the frequency term $f^2$ corresponds to a modification at 5.5 post-Newtonian order, $\Omega_{M}$ is the matter density, and $\Omega_{\Lambda}$ is the dark energy density.
We adopt a $\Lambda$CDM Cosmology with parameters $\Omega_{M}=0.3075$, $\Omega_{\Lambda}=0.691$, and $H_0=67.8~{\rm km~s^{-1} ~Mpc^{-1}}$ \cite{Planck2015} to convert luminosity distance to redshift for \cref{eq:deltahpsi}.  
As most \ac{GW} detections are from the local Universe, we make the simplifying assumption that $c_1,c_2$ are constants and $c_1-c_2$ is of order unity.  This is done by attributing its order of magnitude to $M_\mathrm{PV}$.
Also note that we do not consider the special case where $c_1=c_2$ exactly, and thus $\mu_A=0$, in this work.
This is the case for dynamical Chern-Simons gravity \cite{Yunes:2010yf,cs4,yunes2018} and the constraints on amplitude birefringence in this case can be found in Refs.~\cite{Nair:2019iur, Okounkova:2021xjv}.

\section{Bayesian Inference}
\label{section:PE}

We use Bayesian parameter estimation and model selection to test GW birefringence.
Given data ${d(t)}$, which is a sum of the detector noise $n(t)$ and a possible \ac{GW} signal $h(t,\vec{\theta})$ characterized by parameters $\vec{\theta}$, Bayes theorem states that
\be\label{eq:pe}
P(\vec{\theta}|d,H) = \frac{ P(d|\vec{\theta},H) P(\vec{\theta}|H)} {P(d|H)},
\ee
where $P(\vec{\theta}|d,H)$ is the posterior probability distribution for parameters $\vec{\theta}$, 
$P(\vec{\theta}|H)$ is the prior distribution containing any \textit{a priori} information, 
$P(d | \vec{\theta},H)$ is the likelihood for obtaining the data given model parameters, $P(d|H)$ is a normalization factor called evidence, and
$H$ is the hypothesis for modeling the data.
This work considers two competing hypotheses: $\mathcal{H}_\mathrm{GR}$ where GWs are accurately described by \ac{GR} and $\mathcal{H}_\mathrm{nonGR}$ where GWs have birefringence and are described by Eq.~(\ref{eq:pvwaveform}).
The Bayes factor, or the ratio of the evidences of two hypotheses,
\begin{equation}
\mathcal{B}^\mathrm{nonGR}_\mathrm{GR} = \frac{P(d|\mathcal{H}_\mathrm{nonGR})}{P(d|\mathcal{H}_\mathrm{GR})},
\end{equation} 
quantifies the degree that data favor one hypothesis over another.

Using \texttt{PyCBC Inference} \cite{Biwer_2019}, we numerically sample over all \ac{GW} intrinsic (mass $m_{1,2}$, spin $\vec{s}_{1,2}$, and, in the case of neutron stars, tidal deformability $\Lambda_{1,2}$) and extrinsic parameters (luminosity distance $d_{L}$, inclination angle $\iota$, polarization angle $\Psi$, right ascension $\alpha$, declination $\delta$, coalescence time $t_{c}$, and phase $\phi_c$) as well as the parity violation parameter $M_\mathrm{PV}^{-1}$ for $\mathcal{H}_1$.  
The priors for the intrinsic and extrinsic parameters are the same as those in the 4-OGC \cite{Nitz:2021uxj} which are uniform for mass, spin amplitude, polarization, coalescence phase, and time. The distance prior is uniform in comoving volume.  The spin orientation, sky location, and spatial orientation priors are isotropic.  
The prior for $M_\mathrm{PV}^{-1}$ is uniform in [0,200] GeV$^{-1}$ except for the two outlier events which use [0,1000].

Assuming $M_\mathrm{PV}^{-1}$ is a universal quantity, the $M_\mathrm{PV}^{-1}$ posteriors from the individual \ac{GW} events can be combined to obtain an overall constraint,
\begin{equation}
\label{eq:combined}
p(M_\mathrm{PV}^{-1}|\{d_i\},H_\mathrm{nonGR}) \propto \prod_{i=1}^{N}  p(M_\mathrm{PV}^{-1}|d_i, H_\mathrm{nonGR}),
\end{equation}
where $d_i$ denotes data of the $i$-th \ac{GW} event.

\section{Results}
\label{section:results}
\begin{figure}[ht]
    \centering
    \includegraphics[width=\columnwidth]{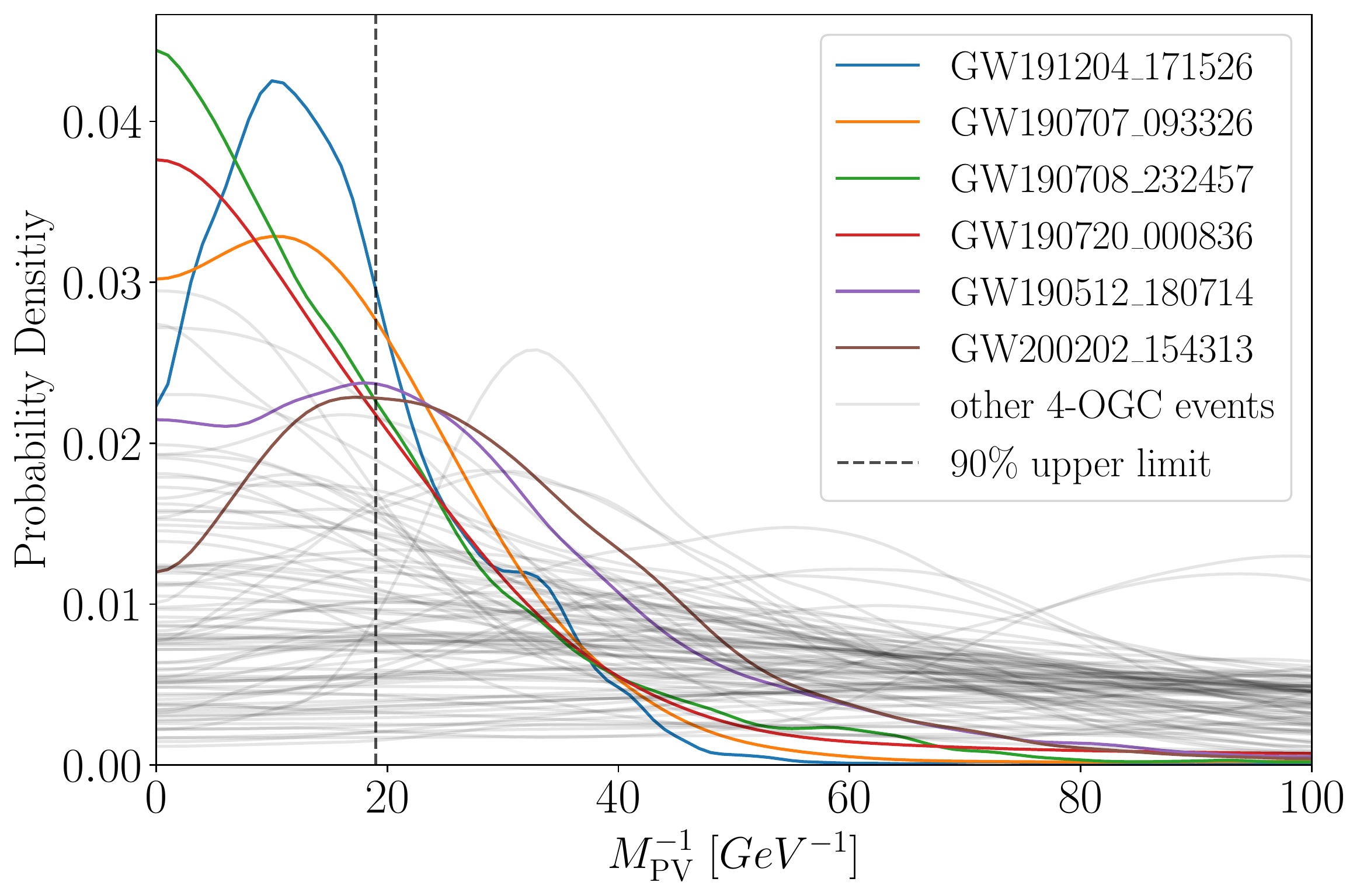}
    \caption{The $M_\mathrm{PV}^{-1}$ posterior distributions for all 4-OGC events except GW190521 and GW191109.
    The legend indicates the events that give the tightest constraints.
    The vertical dashed line denotes the $90\%$ upper limit for $M_\mathrm{PV}^{-1}$ from combined results.}
    \label{fig:mpv}
\end{figure}

We find that ninety-two out of ninety-four events are consistent with \ac{GR}; the $M_\mathrm{PV}^{-1}$ posteriors are shown in \cref{fig:mpv}, with the most constraining events individually indicated in the legend. 
In general, the tightest constraints are all from events with relatively low mass ($\sim10~M_\odot$) binary black hole mergers e.g., GW190707\_093326 has component masses of $12.9M_\odot$ and $7.7M_\odot$ \cite{Nitz:2021uxj}.
This result is unsurprising because the birefringence effect is more significant at higher frequencies, where low mass binaries spend more time (see \cref{eq:deltahpsi}).

The overall constraint is obtained by multiplying the posterior distributions of the individual events together using \cref{eq:combined}.  
We find the $90\%$ upper limit of $M_\mathrm{PV}^{-1}$ to be $19~\mathrm{GeV}^{-1}$, which corresponds to $M_\mathrm{PV} > 0.05$ GeV.
This result is more stringent than previous results, based on twelve \ac{GW} evens, by a factor of five \cite{Wang:2020cub}.  Note that Ref.~\cite{Wang:2020cub} used $h=1$ not $\hbar=1$.  Taking this into account, their result is $M_\mathrm{PV} >0.01$ GeV.
This improvement is due to the increased number of events analyzed and the use of improved \ac{GW} waveforms \cite{Pratten:2020ceb} with longer duration and higher harmonic modes.  

The limit on $M_\mathrm{PV}$ can be easily mapped to bounds on the \ac{SME} coefficients that describe isotropic birefringence of \ac{GW}s at mass dimension $d=5$, via $\big| {\vec{\varsigma}}^{\,(5)} \big| \sim \frac{1}{4} M_\mathrm{PV}^{-1} $~\cite{mewes, Mewes:2019dhj}. 
For $M_\mathrm{PV} > 0.05\,{\rm GeV}$, one gets $\big| {\vec{\varsigma}}^{\,(5)} \big| < 9 \times 10^{-16}\,{\rm m}$, which is comparable to limits from refs.~\cite{Shao:2020shv, Wang:2021ctl}. 

\subsection{GW190521 and GW191109}
\label{subsecton:190521}

\begin{figure}[ht]
    \centering
    \includegraphics[width=\columnwidth]{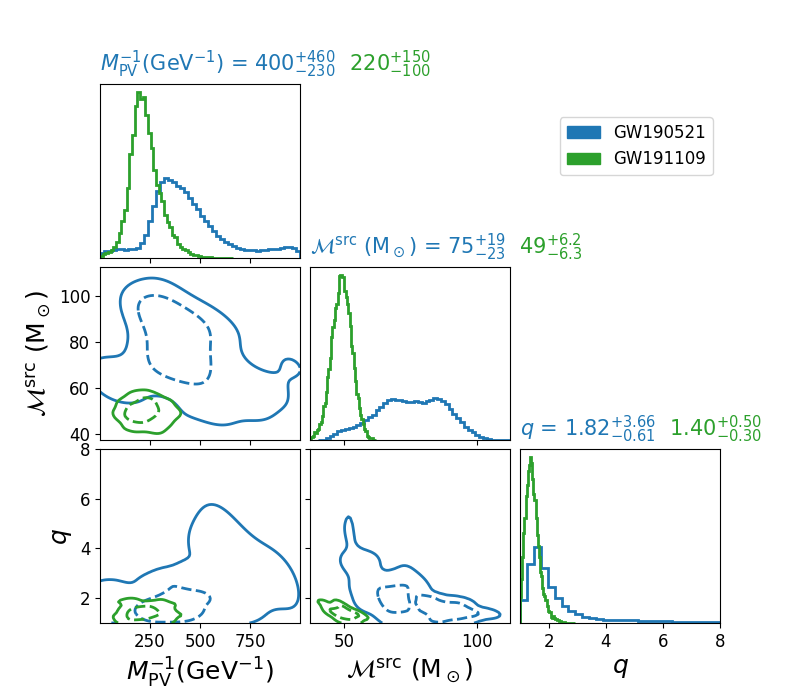}
    \caption{Chirp mass, mass ratio, and $M_\mathrm{PV}^{-1}$ posterior distributions for GW190521 and GW191109.
    The dashed (solid) two-dimensional contours denote the $50\%$ (90\%) credible intervals. 
    The diagonal plots are the one-dimensional marginalization for the posterior.}
    \label{fig:pos}
\end{figure}

We find non-zero results for birefringence in GW190521 and GW191109 with $M_\mathrm{PV}^{-1} = 400^{+460}_{-230}$~GeV$^{-1}$ and $220^{+150}_{-100}$~GeV$^{-1}$ ($90\%$ credible interval).
Furthermore, the Bayes factors support the non-\ac{GR} hypothesis: $\mathcal{B}^\mathrm{nonGR}_\mathrm{GR} = 11.0$ and $22.9$, respectively ($\ln \mathcal{B}^\mathrm{nonGR}_\mathrm{GR} = 2.4$ and $3.1$).
The $M_\mathrm{PV}^{-1}$, source frame chirp mass $\mathcal{M}^{src}=(m_1m_2)^{3/5}/(m_1+m_2)^{1/5}$, and mass ratio $q=m_1/m_2$, posteriors are shown in \cref{fig:pos}. $m_{1/2}$ are the heavier/lighter binary component masses.
Intriguingly, GW190521 and GW191109 are the most and second most massive events found in 4-OGC.

To investigate any systematics causing the apparent deviation from \ac{GR}, we perform two more birefringence tests for the outlier events using a different phenomenological template model IMRPhenomPv3HM \cite{Khan:2019kot} and a numerical relativity surrogate model NRSur7dq4 \cite{PhysRevResearch.1.033015}. 
We find consistent support for non-zero $M_\mathrm{PV}^{-1}$ in the posteriors for these runs.
We further check the data quality around GW190521 and GW191109 by calculating the background noise \ac{PSD} variation {(see \cite{Mozzon:2020gwa} for definition)}, which measures the noise non-stationarity.
For GW190521, the \ac{PSD} variation in a one-hour interval around the event {deviates from Gaussian stationary noise by $\lesssim$ 0.1 (except for a glitch in LIGO Hanford 400 s after the event)}, showing no significant deviation from other ordinary times.
However, the LIGO detector data for GW191109 contains non-Gaussian and non-stationary transient noise artifacts or glitches \cite{GWTC-3}.  
Due to this, we performed our analysis using data with the glitch removed released by LIGO/Virgo \cite{GWTC-3}.

To further quantify whether detector noise could be the cause of the observed non-zero result, we simulate 100 GR signals with parameters drawn from the GW190521 and GW191109 posteriors \cite{Nitz:2021zwj} and inject them into the LIGO/Virgo detector noise nearby the GW190521 and GW191109 triggers, respectively (see the appendix for technical details). We find only 1(4) events, out of 100 injections, have $\ln B^\mathrm{nonGR}_\mathrm{GR}$ larger than what we found for GW190521(GW191109). We thus conclude that the false alarm rate for detecting birefringence is 1(4) in 100 observations.

\begin{figure}[ht]
    \centering
        \includegraphics[width=\columnwidth]{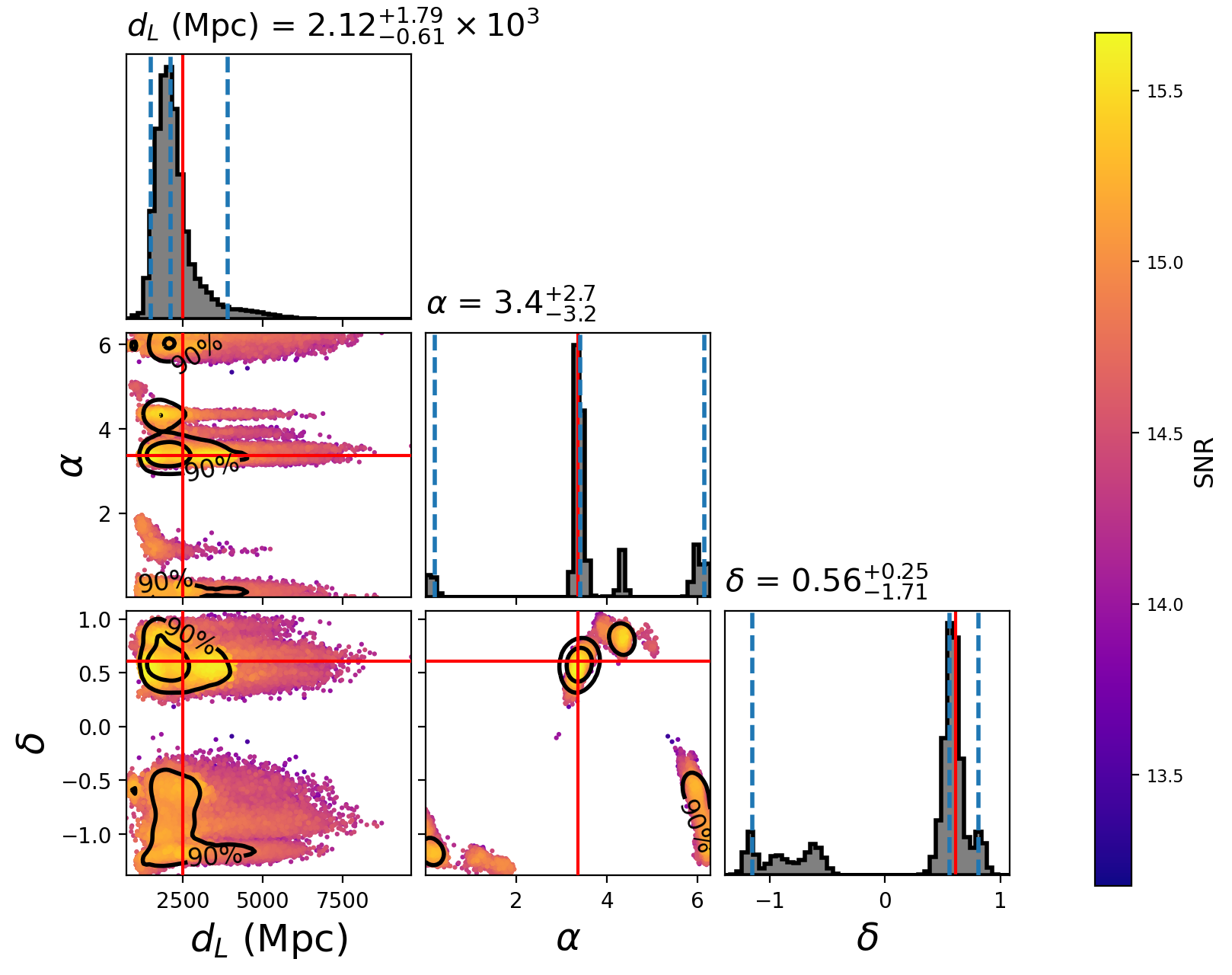}
    \caption{Posterior distributions for luminosity distance $d_L$, right ascension $\alpha$, and declination $\delta$ for GW190521 assuming birefringence.
The median values and $90\%$ credible interval are denoted with dotted vertical lines; the vertical color bar shows the \acl{SNR}. 
    The red lines mark the sky location of a possible EM flare associated with GW190521. }
    \label{fig:gw190521}
\end{figure}

Lastly, we consider the possible \ac{EM} counterpart for GW190521 reported by the ZTF \cite{Graham:2020gwr,Ashton:2020kyr}. Interestingly, we find that including birefringence significantly improves the chance of association.
\cref{fig:gw190521} shows the right ascension ($\alpha$), declination ($\delta$), and luminosity distance posteriors from the birefringence analysis. The red lines mark the independent measurements by ZTF ($\alpha=3.36$ rad, $\delta=0.61$ rad, and redshift=0.438).
The Bayes factor supports coincidence with  $\ln \mathcal{B}_\mathrm{overlap} = 6.6$ in favor of the association.  
The \ac{GR} analyses did not favor association. For instance, Ref.~\cite{Nitz_2021} reports $\ln \mathcal{B}_\mathrm{overlap} \sim -4.4$.

\section{Discussion and Conclusion}
\label{section:conclusion}

We test for \ac{GW} propagation birefringence using state-of-the-art waveform templates and 4-OGC,  the most extensive GW catalog currently available.
Combining the results from all of the events except the outliers GW190521 and GW191109, we constrain the lower energy scale cutoff to $M_\mathrm{PV}> 0.05 ~\mathrm{GeV}$, which is an improvement over previous constraints by a factor of $5$.
The constraint on $M_\mathrm{PV}$ allows us to limit the \ac{SME} isotropic birefringence parameter with mass dimension $d=5$ to $\big| {\vec{\varsigma}}^{\,(5)} \big| < 9 \times 10^{-16}\,{\rm m}$.  These results show that \ac{GW} astronomy is a promising future avenue by which to study gravity at high energies.

We surprisingly find evidence in support of \ac{GW} birefringence for GW190521 and GW191109, which happen to be the most and second most massive events in 4-OGC. 
Furthermore, we find that including birefringences increases the likelihood of association between GW190521 and its possible \ac{EM} counterpart.

We find no disparity between three state-of-the-art waveform approximants and no significant issues with the data quality for GW190521. While there is a glitch in the GW191109 data, it was removed by LIGO/ Virgo, and our analysis suggests the noise fluctuation is unlikely to have caused the non-zero $M^{-1}_\mathrm{PV}$ result.

However, it is well documented that GW190521 is an exceptional event that may not fit well into the simple quasi-circular binary black hole merger picture.
For instance, Refs.~\cite{Romero-Shaw:2020thy, Gayathri:2020coq} show that GW190521 is consistent with the merger of a binary black hole system with eccentric orbit, Ref.~\cite{Gamba:2021gap} gives a Bayes factor that highly favors a hyperbolic encounter over a quasi-circular merger, and Ref.~\cite{PhysRevLett.126.081101} shows that GW190521 could be genuinely new physics, such as a Proca star collision.
The accuracy of current \ac{GR} waveform approximants is limited at the merger stage. This is quite relevant for GW190521 and GW110919 as most of the data is in the merger band.
Our work provides further evidence for non-standard physical effects in \ac{GW} data, which the available \ac{GR} waveform approximants cannot currently account for.
Even if the apparent deviation from \ac{GR} is from new physics, our GW birefringence model can not provide a universal explanation.
A possible extension might include a parity violation that depends on the masses of the binary.

As the advanced LIGO and Virgo detectors are upgraded and the KAGRA detector joins the network, we expect more high mass detections similar to GW190521 and GW191109, which may provide further insight into the physics behind the observed behavior of these outliers.

We release all posterior files and the scripts necessary to reproduce this work in \url{https://github.com/gwastro/4ogc-birefringence}.

\acknowledgments
Y.F.W. and S.M.B. acknowledge the Max Planck Gesellschaft and thank the computing team from AEI Hannover for their significant technical support. 
L.S. was supported by the National Natural Science Foundation of China (11975027, 11991053), the National SKA Program of China (2020SKA0120300), the Young Elite Scientists Sponsorship Program by the China Association for Science and Technology (2018QNRC001), and the Max Planck Partner Group Program funded by the Max Planck Society.

This research has made use of data from the Gravitational Wave Open Science Center (https://www.gw-openscience.org), a service of LIGO Laboratory, the LIGO Scientific Collaboration and the Virgo Collaboration. LIGO is funded by the U.S. National Science Foundation. Virgo is funded by the French Centre National de Recherche Scientifique (CNRS), the Italian Istituto Nazionale della Fisica Nucleare (INFN) and the Dutch Nikhef, with contributions by Polish and Hungarian institutes. 
\vspace{5mm}

\appendix
\section{Investigation on data quality systematics by GR injection} 
We investigate the data quality to determine if non-stationary or non-Gaussian noise could be responsible for the apparent deviation from \ac{GR} for GW190521 and GW191109.
We generate 100 GR waveforms with the IMRPhenomXPHM \cite{Pratten:2020ceb} template and inject them into the LIGO and Virgo data near the coalescence times of GW190521 and GW191109.
The waveform source parameters (component masses and spins, sky location, source orbital orientation, \ac{GW} polarization angle, and coalescence phase) are sampled from the GW190521 and GW191109 posteriors released by 4-OGC \cite{Nitz:2021zwj}.
For GW190521, the waveforms are injected into the time interval [-20, 20] seconds around the trigger time of GW190521. However, we exclude the region [-4,4] seconds around the trigger time, where the event is predominant over the noise. 
Both LIGO detectors contain transient noise artifacts at the merger time of GW191109 \cite{GWTC-3}. We have analyzed GW191109 using a deglitched version of data released by LIGO and Virgo \citep{GWTC-3}. However, a cleaned version of the data in the region of GW191109 is not readily available. To avoid the glitches, which were removed for the GW191109 analysis, we inject our signals into a segment of data  [-100, -30] seconds from the trigger.

The injections are then analyzed the same as the actual event using the method presented in the main text. For the parameter estimation, all priors are the same as those used to analyze the GW190911 and GW190521 respectively.

\begin{figure*}[ht]
    \centering       
     \includegraphics[width=\columnwidth]{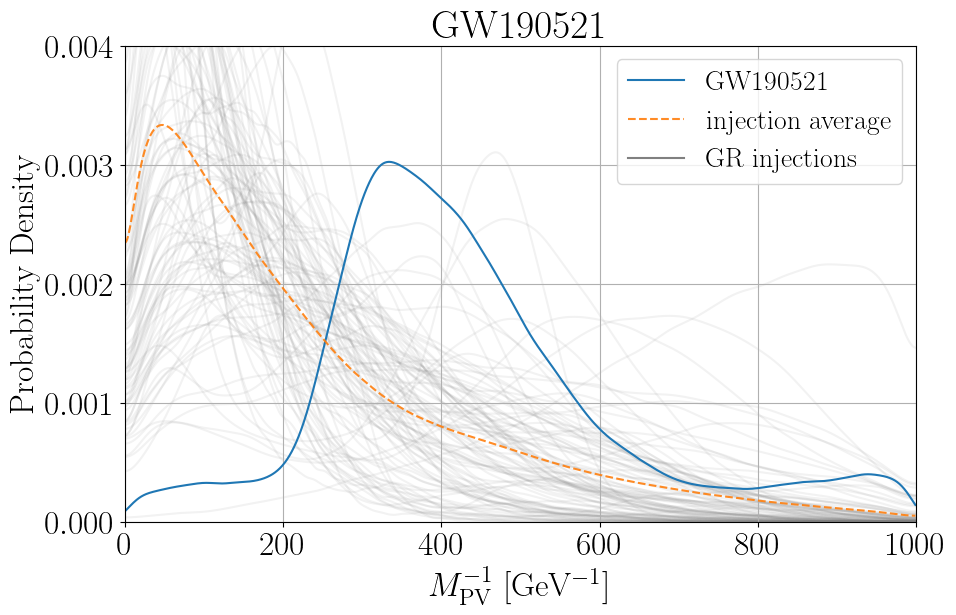}
\includegraphics[width=\columnwidth]{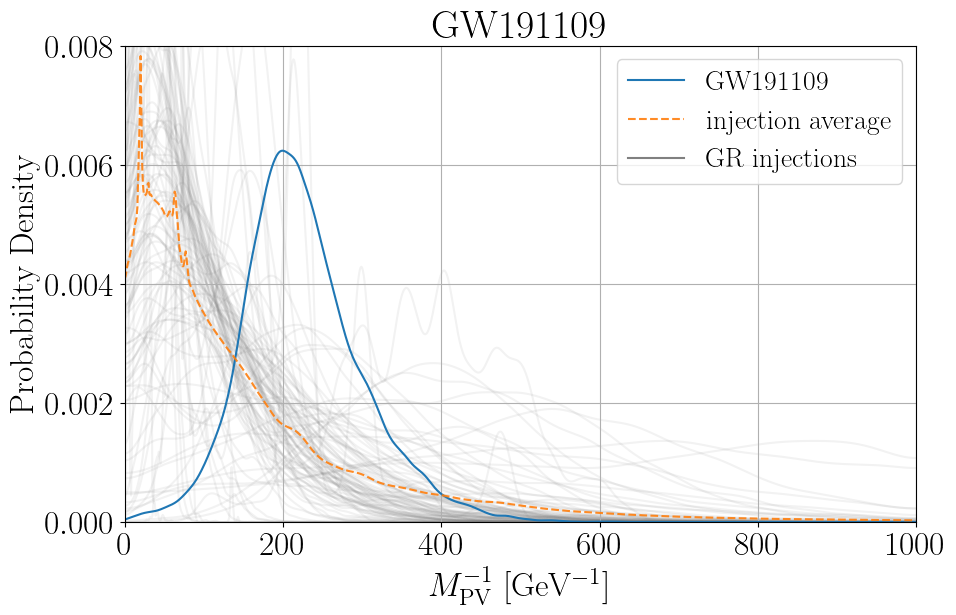}
\includegraphics[width=\columnwidth]{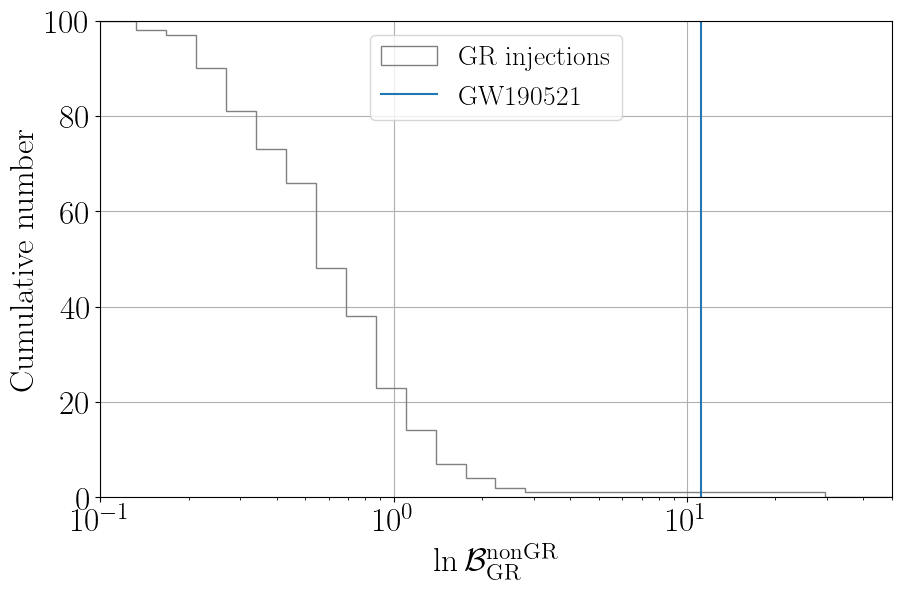}
\includegraphics[width=\columnwidth]{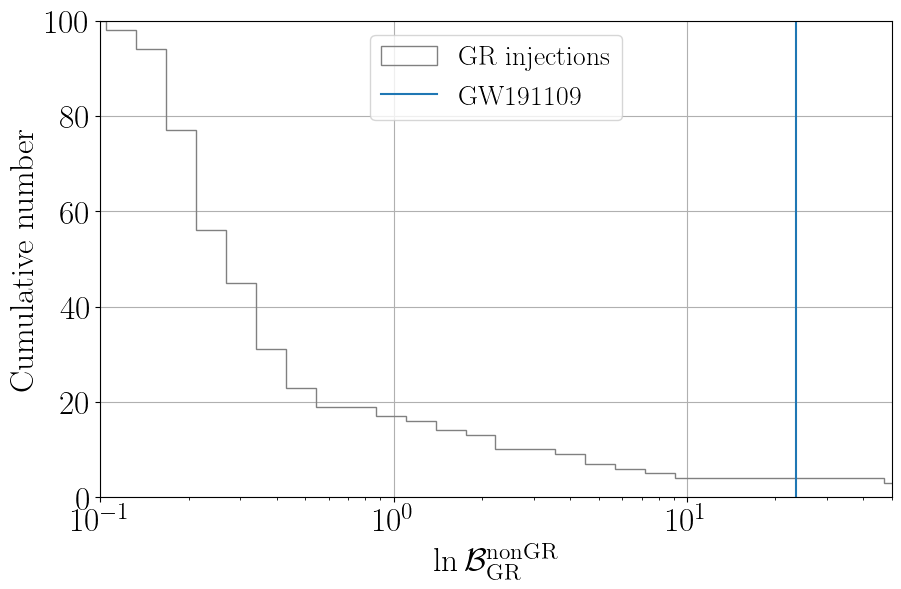}
    \caption{First row: the Bayesian posteriors of $M_\mathrm{PV}^{-1}$ for 100 GR injections that mimic GW190521 (left) and GW191109 (right).
    The posteriors for GW191109 and GW1905121 are plotted for comparison.
    The average posterior of the 100 injections is shown.
    Second row: the cumulative distribution of $\ln \mathcal{B}^\mathrm{nonGR}_\mathrm{GR}$ for GR injections. The histogram shows the number of injections with $\ln \mathcal{B}^\mathrm{nonGR}_\mathrm{GR}$ equal or less than certain values. The Bayes factor for the real events is plotted as a comparison.} 
    \label{fig:sim}
\end{figure*}

The injections' $M_\mathrm{PV}^{-1}$ posteriors are shown in the first row of \cref{fig:sim}. For comparison, the figure also shows the GW190521 and GW191109 posteriors.
We note that some results for GW191109-type injections have a spikey or multi-modal shape; we attribute this to the noisy data around GW191109.

We find that, in rare cases, the background noise fluctuation can produce non-zero peaks for $M_\mathrm{PV}^{-1}$.
To assess the significance of our non-zero $M_\mathrm{PV}^{-1}$ results for GW191109 and GW190521, we extract the Bayes factor from the posteriors using the Savage-Dickey density ratio; the second row of \cref{fig:sim} shows the results.
Using $\ln B^\mathrm{nonGR}_\mathrm{GR}$ as a statistic, only 1 in 100 and 4 in 100 simulations exceed the Bayes factor from GW190521 and GW191109, respectively.
The false alarm rate or p-value to reject the null hypothesis that GR is correct is thus 1(4) out of 100 realizations.

We take the mean of $M_\mathrm{PV}^{-1}$ of all injections to determine the background distribution for the null hypothesis and note that the background is qualitatively different from the actual data results.
Overall we do not find strong evidence that background noise artifacts caused the non-zero results for testing GR.

\bibliography{parity.bib}
\end{document}